\documentclass[a4paper]{article}
\RequirePackage[english]{babel}
\RequirePackage[latin1]{inputenc}
\RequirePackage[T1]{fontenc}
\RequirePackage{mathrsfs}
\RequirePackage{amsmath}
\RequirePackage{amssymb}
\RequirePackage{amsbsy}
\RequirePackage{bm}
\begin{document}
\title{\bf{Weakly-Interacting Massive Particles in Torsionally-Gravitating Dirac Theory}}
\author{Luca Fabbri\\ 
\footnotesize DIME Sez. Metodi e Modelli Matematici, Universit\`{a} di Genova\\
\footnotesize INFN \& Dipartimento di Fisica, Universit\`{a} di Bologna}
\date{}
\maketitle
\begin{abstract}
We shall consider the problem of Dark Matter in torsion gravity with Dirac matter fields; we will consider the fact that if WIMPs in a bath are allowed to form condensates then torsional effects may be relevant even at galactic scales: we show that torsionally-gravitating Dirac fields have interesting properties for the problem of DM. We discuss consequences.
\end{abstract}
\section*{Introduction}
At the present stage, in the race between theories predicting phenomena that experiments must measure and experiments observing facts the theory has to explain, we are in a situation that is quite rare in the history of physics, because although on the one hand there is a vast phenomenology still far from being confirmed on the other hand there are only few things without a proper systematization: among them, one of the most intriguing is certainly Dark Matter.

The problem of Dark Matter consists in the fact that the observed dynamics of the large scale universe, cluster of galaxies and galaxies themselves, seems to be well reproduced by simulations in which the gravitational force is stronger than what is expected to be; this could be due to two factors: a modified theory of gravitation or the same gravity of an exceeding matter distribution.

Of these two approaches, the former might be able to describe some properties like galactic rotation curves, but it can say nothing about other phenomena such as the Bullet Cluster, which is the smallest of a couple of galaxies passing through each other, where during the crossing matter slows down due to the mutual gravitational attraction; however, gravitational lensing has been observed even out of the visible matter distribution, suggesting that there must be an invisible field very weakly-interacting which is nevertheless the source over large distances of a gravitational field: this implies that what causes the gravitational field outside the visible matter distribution cannot be an additional gravitational effect of that matter, because if this were true it would be impossible to have this matter distribution with its own gravitational field in leading-order residing within the visible matter distribution but with higher-order corrections dislodged out of the visible matter distribution itself, and so beside the visible matter another form of matter must be present \cite{Clowe:2006eq}. Thus, it remains the latter approach, describing DM as a real although yet unknown form of matter.

In terms of this approach, DM is a form of matter which must be neutral and very weakly-interacting so to justify why it is invisible and does not suffer the slowing process of the gravitational pull, and there are a few candidates possessing these features: the most relevant are Axions, ELKO and Weakly-Interacting Massive Particles WIMP; we shall briefly discuss them next.

The basic idea of Axions has nothing to do with DM, as they were first postulated to solve problems related to chromodynamics, that is the so-called Peccei-Quinn model; it was only afterward that they have been recognized to have the character DM should have: however, this model in its most natural form is being restricted by observations in experiment such as ADMX and the consequently needed fine-tuning is diminishing its elegance. Both ELKO and WIMP are $\frac{1}{2}$-spin spinors, a form of matter well accepted. The ELKO fields are Majorana spinors solving the mass problem by postulating them to verify second-order derivative field equations \cite{a-g/1,a-g/2,a-l-s/1,a-l-s/2,dr-hs}: these particles are a recent and promising attempt to furnish a candidate for DM, although the fact that they are spinors verifying higher-order derivative field equations may create issues for the torsional self-interaction \cite{fabbri/1,fabbri/2,fabbri/3,fabbri}; WIMP fields are $\frac{1}{2}$-spin fields verifying the Dirac equation, so they are both structurally and dynamically defined in terms of a commonly accepted framework. The WIMP field is a rather natural candidate for DM, but because neutrinos are massless, or at least, even if we believe that the existence of neutrino oscillations must necessarily be described in terms of neutrino masses, the hypothetical neutrino masses are not large enough, then neutrinos cannot be slow and therefore do not match some requisite to be WIMP, so that WIMP fields must be sought in some enlarged forms of the standard model of particle physics. We are not going to discuss here the extensions in which WIMP candidates can be found, since we shall focus on a different type of problem, that is assuming that WIMP can actually be found, then are there interesting properties that ought be investigated?

To be more specific, let us assume that WIMP fields are the correct description of DM: as WIMP fields are $\frac{1}{2}$-spin spinor fields verifying the Dirac equation, then in a gravitational context they are described by the Sciama-Kibble completion of the Einstein theory for the Dirac matter, that is the Sciama-Kibble-Einstein-Dirac SKED theory, where torsional contributions induce fermionic self-interactions in the matter field equation, as it is discussed for example in references \cite{h-h-k-n} and \cite{f/1a,f/1b,f/2a,f/2b,f/3}; these torsionally-induced fermion-fermion interactions can be equivalently rewritten in the form of Nambu-Jona--Lasinio NJL potentials \cite{Baekler:2011jt,Fabbri:2011kq,Fabbri:2012ag}. The consequence of this fact is that WIMP fields permeating the galaxy are described by the SKED theory, therefore subject to a phenomenon of condensation analogous to the one happening in the NJL model; the fact that condensate fields may have a quantum but nevertheless macroscopic structure is expected, and if there is no fundamental interaction such as chromodynamics or electrodynamics confining the WIMP condensate field then it is not unreasonable that the macroscopic scale in this context may even be the galactic scale \cite{Boehmer:2007um,Silverman:2002qx}. As WIMP bath forming a single condensate at galactic scales is the most natural environment in which torsion may be relevant for the galactic rotation curves then we believe that condensates are the most natural systems in which to exploit those torsional effects that have been studied in a classical context in \cite{Tilquin:2011bu}, already with intriguing results.

On the other hand, in the usually accepted description of galactic rotation curves, the orbital velocity of a body within the matter distribution has the Newtonian behaviour linearly increasing with the distance before becoming constant as we move far from the center of the galaxy, which means that the density must scale according to $\frac{1}{r}$ for the visible matter contribution and according to $\frac{1}{r^{2}}$ for the DM contribution; however, thinking at the $\frac{1}{r}$ behaviour as due to the Einsteinian gravitational effects of visible matter and at the $\frac{1}{r^{2}}$ behaviour as due to the Einsteinian gravitational effects of WIMP is unsatisfactory since there is no reason why similar matter fields would have to behave so differently.

Instead, if we think at the $\frac{1}{r}$ behaviour as still due to the Einsteinian gravitational effect of visible matter but at the $\frac{1}{r^{2}}$ behaviour as now due to the torsionally-gravitating contribution of WIMP it is easy to see why they behave differently, and as a consequence we have that the correct behaviour is obtained without the impression of an accidental situation.

In the present paper we will assume this point of view, eventually drawing some of its most relevant consequences.
\section{WIMP Fields in SKED Theory}
As we just mentioned, our starting point is to assume that WIMP fields exists in a galactic context, describing them in terms of the Sciama-Kibble torsional completion of Einstein gravity for Dirac matter fields: for the SKED theory we refer to \cite{f/1a,f/1b,f/2a,f/2b,f/3} and \cite{Fabbri:2011kq,Fabbri:2012ag} for the fundamental definitions and the basic conventions; the formalism is the standard one but because we employ different notation, we will recall anyway some of them for the ease of the reader.

In the paper, we will consider the metric tensors as $g_{\alpha\sigma}$ and $g^{\alpha\sigma}$ with connection $\Gamma^{\alpha}_{\mu\nu}$ defining a covariant derivative $D_{\mu}$ for which $Dg\!=\!0$ and such that torsion tensor defined in terms of $Q^{\alpha}_{\phantom{\alpha}\mu\nu}\!=\!\Gamma^{\alpha}_{[\mu\nu]}$ is taken to be completely antisymmetric \cite{f/1a,f/1b}: the metric-compatibility condition and complete antisymmetry of torsion make the connection decomposable according to the formula
\begin{eqnarray}
&\Gamma^{\mu}_{\sigma\pi}
=\frac{1}{2}Q^{\mu}_{\phantom{\mu}\sigma\pi}
+\frac{1}{2}g^{\mu\rho}\left(\partial_{\pi}g_{\sigma\rho}
+\partial_{\sigma}g_{\pi\rho}-\partial_{\rho}g_{\sigma\pi}\right)
\label{connection}
\end{eqnarray}
while the Riemann curvature tensor is given by
\begin{eqnarray}
&G^{\mu}_{\phantom{\mu}\rho\sigma\pi}=\partial_{\sigma}\Gamma^{\mu}_{\rho\pi}
-\partial_{\pi}\Gamma^{\mu}_{\rho\sigma}
+\Gamma^{\mu}_{\lambda\sigma}\Gamma^{\lambda}_{\rho\pi}
-\Gamma^{\mu}_{\lambda\pi}\Gamma^{\lambda}_{\rho\sigma}
\label{Riemann}
\end{eqnarray}
antisymmetric in the first and second couple of indices, so with one independent contraction $\!G^{\alpha}_{\phantom{\alpha}\rho\alpha\sigma}\!\!=\!G_{\rho\sigma}$ with $\!G_{\rho\sigma}g^{\rho\sigma}\!\!=\!G$ called Ricci tensor and scalar, and
\begin{eqnarray}
&G^{\mu}_{\phantom{\mu}\rho\sigma\pi}\!=\!R^{\mu}_{\phantom{\mu}\rho\sigma\pi}
\!+\!\frac{1}{2}(\nabla_{\sigma}Q^{\mu}_{\phantom{\mu}\rho\pi}
-\nabla_{\pi}Q^{\mu}_{\phantom{\mu}\rho\sigma})
\!+\!\frac{1}{4}(Q^{\mu}_{\phantom{\mu}\lambda\sigma}Q^{\lambda}_{\phantom{\lambda}\rho\pi}
-Q^{\mu}_{\phantom{\mu}\lambda\pi}Q^{\lambda}_{\phantom{\lambda}\rho\sigma})
\label{decomposition}
\end{eqnarray}
in terms of the torsionless covariant derivative $\nabla_{\sigma}$ and torsionless curvature given by $R^{\mu}_{\phantom{\mu}\rho\sigma\pi}$ such that $R^{\alpha}_{\phantom{\alpha}\rho\alpha\sigma}\!\!=\!R_{\rho\sigma}$ and $R_{\rho\sigma}g^{\rho\sigma}\!\!=\!R$ as usual; the coordinate formalism can be translated in the tetrad formalism upon definition of the dual bases of orthonormal tetrads $\xi^{a}_{\sigma}$ and $\xi_{a}^{\sigma}$ such that they verify orthonormality conditions given by $\xi_{a}^{\sigma}\xi_{b}^{\nu}g_{\sigma\nu}\!=\!\eta_{ab}$ and $\xi^{a}_{\sigma}\xi^{b}_{\nu}g^{\sigma\nu}\!=\!\eta^{ab}$ in terms of the Minkowskian matrices, while the spin-connection $\Gamma^{i}_{j\mu}$ defining the covariant derivative $D_{\mu}$ is such that it gives $D\xi\!=\!0$ and $D\eta\!=\!0$ and for a connection with two different types of indices one cannot define torsion: these conditions imply that (\ref{connection}) is
\begin{eqnarray}
&\Gamma^{b}_{\phantom{b}j\mu}=
\xi^{\alpha}_{j}\xi_{\rho}^{b}\left(\Gamma^{\rho}_{\phantom{\rho}\alpha\mu}
+\xi_{\alpha}^{k}\partial_{\mu}\xi^{\rho}_{k}\right)
\label{spin-connection}
\end{eqnarray}
and it is antisymmetric in the two world indices while the curvature is
\begin{eqnarray}
&G^{a}_{\phantom{a}b\sigma\pi}
=\partial_{\sigma}\Gamma^{a}_{b\pi}-\partial_{\pi}\Gamma^{a}_{b\sigma}
+\Gamma^{a}_{j\sigma}\Gamma^{j}_{b\pi}-\Gamma^{a}_{j\pi}\Gamma^{j}_{b\sigma}
\label{Riemanngauge}
\end{eqnarray}
antisymmetric in both the coordinate and the world indices and writable in terms of the Riemann curvature as $G^{ab}_{\phantom{ab}\sigma\pi}\!=\!G^{\mu\nu}_{\phantom{\mu\nu}\sigma\pi} \xi^{a}_{\mu}\xi^{b}_{\nu}$ as obvious. The advantage of such change of formalism is that the most general coordinate transformations of the coordinate formalism (with Greek indices) are equivalently written in terms of the special Lorentz transformations of the tetrad formalism (with Latin indices) which admits a specific representation, suitable of being the usual real one but also a new complex one; in the tetrad formalism then, complex representations are definable and therefore we may proceed to the introduction of complex Lorentz transformations. These are called spinorial transformations.

In such a geometrical background, spinor fields will be taken to be the simplest $\frac{1}{2}$-spin spinors, defined in terms of the $2$-dimensional sigma matrices $\vec{\boldsymbol{\sigma}}$  so that the most general Lorentz complex transformation can be written according to the expressions $\mathrm{exp}[(\vec{\varphi}\!+\!i\vec{\theta})\!\cdot\!\vec{\frac{\boldsymbol{\sigma}}{2}}]$ or $\mathrm{exp}[(-\vec{\varphi}\!+\!i\vec{\theta})\!\cdot\!\vec{\frac{\boldsymbol{\sigma}}{2}}]$ because of the sign ambiguity of the boosts: these can be merged into the reducible $4$-dimensional representation after introducing the $\boldsymbol{\gamma}^{\mu}$ matrices in chiral representation
\begin{eqnarray}
\vec{\boldsymbol{\gamma}}
=\left(\begin{array}{cc}
\boldsymbol{0} & \vec{\boldsymbol{\sigma}}\\
-\vec{\boldsymbol{\sigma}} & \boldsymbol{0}
\end{array}\right)\ \ \ \ \ \ \ \ 
\boldsymbol{\gamma}^{0}
=\left(\begin{array}{cc}
\boldsymbol{0} & \boldsymbol{\mathbb{I}}\\
\boldsymbol{\mathbb{I}} & \boldsymbol{0}
\end{array}\right)
\end{eqnarray}
with sigma matrices $\frac{1}{4}[\boldsymbol{\gamma}_{i},\boldsymbol{\gamma}_{j}]=\boldsymbol{\sigma}_{ij}$ so that $\{\boldsymbol{\gamma}_{i},\boldsymbol{\sigma}_{jk}\}=i\varepsilon_{ijkq}\boldsymbol{\gamma}\boldsymbol{\gamma}^{q}$ in the complete Lorentz complex representation $\mathrm{exp}[\frac{1}{2}\theta^{ij}\boldsymbol{\sigma}_{ij}]$ as it is well-known, that is in the sought spinorial transformation in terms of which the $\frac{1}{2}$-spin spinors will be defined on a general spacetime background. Then it is possible to introduce the spinor-connection $\boldsymbol{A}_{\mu}$ defining the spinor-covariant derivative $\boldsymbol{D}_{\mu}$ containing the information about the dynamics of the spinor fields and for which the spinorial constancy of $\boldsymbol{\gamma}_{j}$ is automatic: the spinor-connection $\boldsymbol{A}_{\mu}$ is given by
\begin{eqnarray}
&\boldsymbol{A}_{\mu}=\frac{1}{2}\Gamma^{ab}_{\phantom{ab}\mu}\boldsymbol{\sigma}_{ab}
\label{spinor-connection}
\end{eqnarray}
in terms of the complex-valued spin-connection and the curvature is given by
\begin{eqnarray}
&\boldsymbol{F}_{\sigma\pi}
=\partial_{\sigma}\boldsymbol{A}_{\pi}-\partial_{\pi}\boldsymbol{A}_{\sigma}
+[\boldsymbol{A}_{\sigma},\boldsymbol{A}_{\pi}]
\label{RiemannMaxwellgauge}
\end{eqnarray}
which is a tensorial spinor antisymmetric in the tensorial indices writable as
\begin{eqnarray}
&\boldsymbol{F}_{\sigma\pi}=\frac{1}{2}G^{ab}_{\phantom{ab}\sigma\pi}\boldsymbol{\sigma}_{ab}
\label{combination}
\end{eqnarray}
in terms of the curvature of the spacetime, in a very compact form.

This defines the basic formalism we are going to employ, in terms of which the kinematic background is now set up, and next point that needs to be settled is the implementation of the dynamics by requiring a link between the geometric fields on the one hand and the material quantities on the other hand by defining the fundamental Lagrangian: as it has been discussed in \cite{Fabbri:2011kq} when we develop the Lagrangian formalism we usually employ a geometric Lagrangian built on the torsional completion of the Ricci scalar, but this only includes torsion implicitly through the connection within the curvature while torsion in general should also be included explicitly in the action itself; since at the least-order derivative in the action the curvature appears linearly and torsion is squared, and because according to our restriction of having a completely antisymmetric torsion there is only one possible squared torsion term, consequently we have that the most general completely antisymmetric torsion completion of the gravitational least-order derivative dynamical action is given according to the following Lagrangian density $\mathscr{L}\!=\!\frac{a-16\pi k}{4a16\pi k}Q^{2}\!+\!\frac{1}{16\pi k}G \!\equiv\!-\frac{1}{4a}Q^{2}\!+\!\frac{1}{16\pi k}R$ in terms of the gravitational constant $k$ and an additional torsional coupling constant that is in general different from the gravitational constant, and which can only be determined empirically. By varying this geometrical Lagrangian we get the system of field equations for the geometry coupling the completely antisymmetric torsion and curvature to the material quantities according to the expressions
\begin{eqnarray}
&Q^{\rho\mu\nu}=-aS^{\rho\mu\nu}
\label{torsion-spin}\\
&\left(\frac{8\pi k}{a}\!\!-\!\!\frac{1}{2}\right)\!\left(\frac{1}{4}\delta^{\mu}_{\nu}Q^{2}
\!-\!\frac{1}{2}Q^{\mu\alpha\sigma}Q_{\nu\alpha\sigma}
\!+\!D_{\rho}Q^{\rho\mu}_{\phantom{\rho\mu}\nu}\right)
\!\!+\!\!\left(G^{\mu}_{\phantom{\mu}\nu}\!-\!\frac{1}{2}\delta^{\mu}_{\nu}G\right)
=8\pi kT^{\mu}_{\phantom{\mu}\nu}
\label{curvature-energy}
\end{eqnarray}
with completely antisymmetric spin $S^{\rho\mu\nu}$ and energy $T^{\mu\nu}$ verifying the set of conservation laws given as usual by the following relationships
\begin{eqnarray}
&D_{\rho}S^{\rho\mu\nu}+\frac{1}{2}\left(T^{\mu\nu}-T^{\nu\mu}\right)\equiv0
\label{conservationspin}\\
&D_{\mu}T^{\mu\nu}
+T_{\rho\beta}Q^{\rho\beta\nu}-S_{\mu\rho\beta}G^{\mu\rho\beta\nu}\equiv0
\label{conservationenergy}
\end{eqnarray}
which are such whenever the matter fields satisfy matter field equations.

The material Lagrangian is given by the Dirac matter field Lagrangian as it is usually done; by complementing the geometrical Lagrangian with the Dirac matter field Lagrangian, the variation with respect to the matter field gives
\begin{eqnarray}
&S^{\rho\mu\nu}=\frac{i\hbar}{4}\overline{\psi}\{\boldsymbol{\gamma}^{\rho},\boldsymbol{\sigma}^{\mu\nu}\}\psi
\label{spin}\\
&T^{\mu}_{\phantom{\mu}\nu}
=\frac{i\hbar}{2}\left(\overline{\psi}\boldsymbol{\gamma}^{\mu}\boldsymbol{D}_{\nu}\psi
-\boldsymbol{D}_{\nu}\overline{\psi}\boldsymbol{\gamma}^{\mu}\psi\right)
\label{energy}
\end{eqnarray}
where the spin is completely antisymmetric and the energy is non-symmetric, and such that they verify the above conservation laws whenever
\begin{eqnarray}
&i\hbar\boldsymbol{\gamma}^{\mu}\boldsymbol{D}_{\mu}\psi-m\psi=0
\label{matterequations}
\end{eqnarray}
are satisfied as matter field equations. Finally, when taken all together we have that the entire system of field equations is given by the equations
\begin{eqnarray}
&Q^{\rho\mu\nu}
=-a\frac{i\hbar}{4}\overline{\psi}\{\boldsymbol{\gamma}^{\rho},\boldsymbol{\sigma}^{\mu\nu}\}\psi
\label{Sciama--Kibble}\\
\nonumber
&\left(\frac{8\pi k}{a}\!\!-\!\!\frac{1}{2}\right)\!\left(\frac{1}{4}\delta^{\mu}_{\nu}Q^{2}
\!-\!\frac{1}{2}Q^{\mu\alpha\sigma}Q_{\nu\alpha\sigma}
\!+\!D_{\rho}Q^{\rho\mu}_{\phantom{\rho\mu}\nu}\right)
\!\!+\!\!\left(G^{\mu}_{\phantom{\mu}\nu}\!-\!\frac{1}{2}\delta^{\mu}_{\nu}G\right)\\
&=8\pi k\frac{i\hbar}{2}\left(\overline{\psi}\boldsymbol{\gamma}^{\mu}\boldsymbol{D}_{\nu}\psi
-\boldsymbol{D}_{\nu}\overline{\psi}\boldsymbol{\gamma}^{\mu}\psi\right)
\label{Einstein}
\end{eqnarray}
and the matter field equations above given by
\begin{eqnarray}
&i\hbar\boldsymbol{\gamma}^{\mu}\boldsymbol{D}_{\mu}\psi-m\psi=0
\label{Dirac}
\end{eqnarray}
as a direct calculation would show straightforwardly.

Finally, it is worth noticing that our initial assumption of a completely antisymmetric torsion restrains the description to a completely antisymmetric spin allowing only the simplest spinor field to be defined without constraints, or equivalently, that such a restriction does not constitutes any loss of generality since we are interested in the simplest spinor field alone \cite{f/2a,f/2b}; thus this is the most general system of field equations we may have under the initial conditions with which we want to work, and so these are the field equations we will employ next: in this system of field equations, torsional quantities can be decomposed in terms of torsionless quantities and torsional contributions that can be converted through the torsion-spin coupling field equation (\ref{Sciama--Kibble}) into spinorial potentials, so that the curvature-energy coupling field equations reduce to the form
\begin{eqnarray}
\nonumber
&R_{\mu\nu}=-8\pi k\frac{m}{2}\overline{\psi}\psi g_{\mu\nu}+\\
&+8\pi k\frac{i\hbar}{4}\left(\overline{\psi}\boldsymbol{\gamma}_{\mu}\boldsymbol{\nabla}_{\nu}\psi
+\overline{\psi}\boldsymbol{\gamma}_{\nu}\boldsymbol{\nabla}_{\mu}\psi
-\boldsymbol{\nabla}_{\nu}\overline{\psi}\boldsymbol{\gamma}_{\mu}\psi
-\boldsymbol{\nabla}_{\mu}\overline{\psi}\boldsymbol{\gamma}_{\nu}\psi\right)
\label{gravitation}
\end{eqnarray}
and the matter field equations after a Fierz rearrangement become
\begin{eqnarray}
&i\hbar\boldsymbol{\gamma}^{\mu}\boldsymbol{\nabla}_{\mu}\psi
-\frac{3a}{16}\hbar^{2}\left(\overline{\psi}\psi\mathbb{I}
-\overline{\psi}\boldsymbol{\gamma}\psi\boldsymbol{\gamma}\right)\psi-m\psi=0
\label{matter}
\end{eqnarray}
where the gravitational field equations for the Ricci tensor are those we would have had in the torsionless case and the matter field equations are those we would have had if there were no torsion but Nambu-Jona--Lasinio potentials.
\subsection{Particle Condensate with Gravitational Corrections}
We may proceed by specifying to the physical situation we want to study, namely that of WIMP fields forming a condensate over galactic distances: that quantum particles in the non-relativistic limit may condensate thus behaving as a single macroscopic field is known, as reviewed for example in \cite{f/3,Fabbri:2012ag}, and the idea that such macroscopic field may stretch to galactic scales has already been put forward, as it may be seen for instance in \cite{Boehmer:2007um,Silverman:2002qx}; the idea is that a bath of quantum particles would be a condensate of entangled entities behaving as a single macroscopic field filling galactic spaces. So the matter field we have will be interpreted as describing the particle condensate seen as a single macroscopic field with the extension of the galactic halo; in this article we will not discuss how this may occur but we will take it for given, developing its consequences for the galactic rotation curves. We will see that it will be possible with no further assumption to reproduce the observed behaviour of the rotating galaxy.

As a first element, we have to specify the physical situation we will consider, that is the case of rotating galaxies: we will consider the case of stationary spherical symmetry, for which, in the frame centered in the center of the galaxy and with equatorial plane coincident with the rotation plane of the galaxy, described with coordinates $(t,r,\theta,\varphi)$, it is widely known that the most general metric is given by $ds^{2}\!=\!A^2dt^{2}\!-\![B^2dr^{2}\!+\!r^{2}(d\theta^{2}\!+\!(\sin{\theta}d\varphi)^{2})]$ in terms of two functions of the radial coordinate $A(r)$ and $B(r)$ to be determined; however in the weak-gravity slow-speed approximation used to study galaxies, it is also known that we may take $A\!\approx\!\frac{1}{B}\!\approx\!1\!+\!2V$ yielding for the curvature the approximated expression $R^{t}_{t}\approx R_{tt}\approx\mathrm{div}\ \mathrm{grad}V$ and we recall that the geodesic equation of motion $\vec{a}+\mathrm{grad}V\!\approx\!0$ gives the acceleration felt by a test-body moving in the gravitational field. In this limit we have $i\overline{\psi}\boldsymbol{\gamma}\psi\approx0$ and so
\begin{eqnarray}
&i\hbar\boldsymbol{\gamma}^{t}\boldsymbol{\nabla}_{t}\psi
+i\hbar\vec{\boldsymbol{\gamma}}\cdot\vec{\boldsymbol{\nabla}}\psi
-\frac{3a}{16}\hbar^{2}\overline{\psi}\psi\psi-m\psi\approx0
\end{eqnarray}
with which temporal derivatives of the spinor are substituted with spatial derivatives of the spinor in the time-time component of the gravitational field equations
\begin{eqnarray}
\nonumber
&\mathrm{div}\vec{a}\!\approx\!-\mathrm{div}\ \mathrm{grad}V\!\approx\!-R_{tt}
\!\approx\!4\pi k\left[m\overline{\psi}\psi
-i\hbar\left(\overline{\psi}\boldsymbol{\gamma}^{t}\boldsymbol{\nabla}_{t}\psi
-\boldsymbol{\nabla}_{t}\overline{\psi}\boldsymbol{\gamma}^{t}\psi\right)\right]\!\approx\!\\
&\!\approx\!-4\pi k\!\left[m\overline{\psi}\psi
+\frac{3}{8}a\hbar^{2}\overline{\psi}\psi\overline{\psi}\psi
+i\hbar\left(\vec{\boldsymbol{\nabla}}\overline{\psi}\cdot\vec{\boldsymbol{\gamma}}\psi
-\overline{\psi}\vec{\boldsymbol{\gamma}}\cdot\vec{\boldsymbol{\nabla}}\psi\right)\right]
\end{eqnarray}
so to remove the entire explicit temporal dependence, with $\vec{\boldsymbol{\gamma}}\!\cdot\!\vec{\boldsymbol{\nabla}}\psi$ denoting the spatial projection of the scalar product between the gamma matrices and the spinorial covariant derivatives: in slow-speed approximation, the stationary configurations of energy $E$ verify $E^{2}\!-\!m^{2}\!\approx\!2m(E\!-\!m)$ and with the spinor in standard representation $\overline{\psi}\!=\!(\phi^{\dagger},-\chi^{\dagger})$ the matter field equation splits as
\begin{eqnarray}
\nonumber
&i\hbar\vec{\boldsymbol{\sigma}}\cdot\vec{\boldsymbol{\nabla}}\chi
+\left[E-m-\frac{3a}{16}\hbar^{2}(\phi^{\dagger}\phi-\chi^{\dagger}\chi)\right]\phi\approx0\\
&i\hbar\vec{\boldsymbol{\sigma}}\cdot\vec{\boldsymbol{\nabla}}\phi
+\left[E+m+\frac{3a}{16}\hbar^{2}(\phi^{\dagger}\phi-\chi^{\dagger}\chi)\right]\chi\approx0
\end{eqnarray}
and the gravitational field equation is
\begin{eqnarray}
\nonumber
&\mathrm{div}\vec{a}\!\approx\!-4\pi k[m(\phi^{\dagger}\phi-\chi^{\dagger}\chi)
+\frac{3}{8}a\hbar^{2}(\phi^{\dagger}\phi-\chi^{\dagger}\chi)(\phi^{\dagger}\phi-\chi^{\dagger}\chi)+\\
&+i\hbar(\vec{\boldsymbol{\nabla}}\phi^{\dagger}\cdot\vec{\boldsymbol{\sigma}}\chi
-\phi^{\dagger}\vec{\boldsymbol{\sigma}}\cdot\vec{\boldsymbol{\nabla}}\chi
+\vec{\boldsymbol{\nabla}}\chi^{\dagger}\cdot\vec{\boldsymbol{\sigma}}\phi
-\chi^{\dagger}\vec{\boldsymbol{\sigma}}\cdot\vec{\boldsymbol{\nabla}}\phi)]
\end{eqnarray}
which is the Newton law we are seeking. The two matter field equations in their semi-spinorial form can be plugged into one another, after which one sees that the semi-spinor $\chi$ tends to vanish, justifying its usual name of \emph{small} semi-spinor, while the semi-spinor $\phi$ is still present satisfying the field equation given by
\begin{eqnarray}
&\frac{\hbar^{2}}{2m}\boldsymbol{\nabla}^{2}\phi
-\frac{9\hbar^{4}a^{2}}{512m}\phi^{\dagger}\phi\phi^{\dagger}\phi\phi
-\frac{3\hbar^{2}a}{16}\phi^{\dagger}\phi\phi+(E-m)\phi\approx0
\end{eqnarray}
justifying its usual name of \emph{large} semi-spinor satisfying the Schr\"{o}dinger field equation, and so we also have that the term with spatial spinorial derivatives vanishes in the gravitational field equation leaving the simpler form
\begin{eqnarray}
&\mathrm{div}\vec{a}\approx-4\pi k(m\phi^{\dagger}\phi
+\frac{3}{8}a\hbar^{2}\phi^{\dagger}\phi\phi^{\dagger}\phi)
\end{eqnarray}
with algebraic contributions of the large semi-spinorial component alone.

Notice that the absence of any magnetic field allows us to consider the large semi-spinor's spin-up and spin-down projections as independent, so that we lose no generality in taking the semi-spinor field as $\phi^{\dagger}\!=\!(u^{\ast},0)$ with $u$ verifying
\begin{eqnarray}
&\frac{\hbar^{2}}{2m}\boldsymbol{\nabla}^{2}u
-\frac{9\hbar^{4}a^{2}}{512m}u^{5}-\frac{3\hbar^{2}a}{16}u^{3}+(E-m)u\approx0
\label{Schroedinger}
\end{eqnarray}
and with
\begin{eqnarray}
&\mathrm{div}\vec{a}\approx-4\pi k(mu^{2}+\frac{3}{8}a\hbar^{2}u^{4})
\label{Newton}
\end{eqnarray}
which now have to be solved, and this is what we are trying to do next.
\subsubsection{Large-density solutions for constant-valued curves}
So far we have obtained the field equations with which we will work, that is the Schr\"{o}dinger equation (\ref{Schroedinger}) and the Newton law (\ref{Newton}): we must solve the Schr\"{o}dinger equation, plugging the solution into the Newton law as to see what are the corrections to the galactic rotation curves in parallel to \cite{Tilquin:2011bu}; one point we need to remember is that the field is thought to represent condensates, which has high-density field distributions. In fact it is the high-density field distribution what makes relevant the non-linear potentials; these non-linear potentials can be as relevant or even more relevant than the linear term. We will see what happens to the field equations if the highest-order term overtakes all others.

To proceed to the calculation, let us first take into account the Schr\"{o}dinger equation and the Newton law written in stationary spherically symmetric coordinates and with the above high-density field distribution condition according to which we retain only the largest-power potential: according to such a condition the Schr\"{o}dinger field equation is consequently given by the expression
\begin{eqnarray}
&\frac{1}{r^{2}}\left[\frac{\partial}{\partial r}\left(r^{2}\frac{\partial u}{\partial r}\right)
+\frac{1}{\sin{\theta}}\frac{\partial }{\partial \theta}
\left(\sin{\theta}\frac{\partial u}{\partial \theta}\right)\right]
-\frac{9\hbar^{2}a^{2}}{256}u^{5}\approx0
\label{Sapp}
\end{eqnarray}
while the Newton law is given by
\begin{eqnarray}
&\frac{1}{r^{2}}\frac{\partial}{\partial r}\left(r^{2}a\right)
\approx\frac{3}{2}\pi ka\hbar^{2}u^{4}
\label{Napp}
\end{eqnarray}
for an approximately circular Keplerian orbit; this form of the Schr\"{o}dinger equation has for possible solution the one given in the form
\begin{eqnarray}
&u=\sqrt{\frac{8}{3\hbar a r\sin{\theta}}}
\label{solution}
\end{eqnarray}
which is square-integrable in the origin although not square-integrable at infinity, but since the condition of high-density field distribution forbids us to reach regions too far away then we are allowed not to care about regions too far outside the galactic halo. Inside the galactic halo such a solution is valid, and we plug it into the Newton law obtaining the following expression
\begin{eqnarray}
&\frac{\partial}{\partial r}\left(r^{2}a\right)
\approx\frac{32\pi k}{3a}\left(\frac{1}{\sin{\theta}}\right)^2
\end{eqnarray}
which has to be solved: writing the centrifugal acceleration in terms of the tangential velocity and taking for simplicity the equatorial plane, we get
\begin{eqnarray}
&v^{2}\approx\frac{32\pi k}{3a}
\end{eqnarray}
spelling that the tangential velocity is nearly constant, with constant value no longer containing any reference to universal constants apart from the purely geometrical ones. Notice that if the torsional constant is taken to be about the same as the Newton constant then this velocity approaches the speed of light, but for larger values of the torsional constant the velocity becomes smaller and for instance if it is about $10^{8}$ times the Newton constant the tangential velocity's constant value becomes about $10^{-3}$ times the speed of light, as measured.

Now that we reached a result it is necessary to go back as to reconsider our assumptions and check their consistency with the result: the assumption that in the Schr\"{o}dinger and Newton field equations (\ref{Schroedinger}-\ref{Newton}) we retained only the largest-density contributions so to get the Schr\"{o}dinger and Newton approximated field equations (\ref{Sapp}-\ref{Napp}) is condensed into $a\hbar^{2}u^{2}\gg m$ as a condition whose validity imposes the mass to be smaller than $10^{-54}$ in Planck units; such a value is ridiculously small but it is not in contradiction with any known physical fact, since on the one hand light particles can energetically be produced easily but on the other hand WIMP particles are defined to have a low scattering amplitude allowing them to escape detection easily as well. Usually torsion is neglected and thus it is necessary to have a mass larger by $34$ orders of magnitude for the WIMP particle bath to ensure the same physical effects: this enormous discrepancy in mass, while still retaining a similarly low capacity to interact with ordinary matter, should make the two types of WIMP particles quite easily distinguishable, so soon as experiments capable of measuring the mass of the WIMP candidate will be devised. When such experiments will be conceived they will be able to immediately rule out the present model if the masses are measured to be considerably larger than the presented upper limits, which means that this approach is clearly falsifiable and thus scientifically reliable.

That the torsional coupling constant value $a$ is in fact about $10^{8}$ times the Newton constant may be much more difficult to check, because such a value has the property of being not too much larger then the Newton constant itself, which avoids a fine-tuning involving too many orders of magnitude, but at the same time this means that it is much smaller compared to the coupling constants of all other interactions, so that the torsionally-induced spin-contact interaction becomes relevant in particle physics much beyond the scale of the nuclear forces, and therefore beyond the possibility to measure it through scattering of elementary particles; this is a weird situation because in order to estimate the precise value of the torsional coupling constant one needs to test either the very large scale gravitational behaviour in cosmology or the very small scale scattering amplitude in particle physics, and while the former case is just reached in the case of Dark Matter the latter case is for the moment beyond the limits of the present accelerators. What this means is that for now we only have a single model to use the torsional interaction with such a value of the torsional coupling constant, and it is impossible to assess how likely this value really is by only studying a single physical effect.
\section*{Conclusions}
So let us summarize what we have done in the present paper: if we were to distill all our hypotheses and assumptions we would find that: first of all, we have considered the hypothesis of existence of WIMP fields filling the underlying geometric background constituted by the most general torsional completion of gravity, that is the one in which we have that the spin-torsion and energy-curvature couplings have place in terms of different coupling constants, according to the most general SKED theory; secondly, we have assumed that the WIMP could form condensates retaining the torsionally-induced non-linear potentials as the most relevant ones, and that the weak-gravity slow-speed approximations were valid as it is usually done in studying galactic dynamics. On these bases, we merely have logically derived all the possible consequences: we have seen that the resulting dynamics of the WIMP condensate gives rise to gravitational corrections for which the galactic rotation curves have a tangential velocity that is shown to be nearly constant, as it is expected for Dark Matter. Because the value of the torsional constant can only be fixed empirically, and it has never been fixed so far, then we ignore what its value could actually be, but for a torsional constant of about $10^{8}$ times the Newton constant the tangential velocity is not only constant but it also has the value it should have to match observations.

It is important to stress it once more: all of the hypotheses and assumptions from which we decided to start, that is existence of WIMP, their possibility to condensate and the condition of having large torsional contributions compatible with a torsional constant of $10^{8}$ in Planck units, are either accepted or seen as reasonable, as discussed in \cite{Boehmer:2007um, Silverman:2002qx} and \cite{Tilquin:2011bu}; also the methods of calculations that we have been employing here are those already commonly used in the known literature. There is no point, being it a principle or a computation, where we considered something never considered before: the novelty of our paper is that these principles have been considered together. And that these principles can stay together consistently is proven by the fact that our results provide a model for galactic rotation curves that fits observations adequately.

The detailed description that comes out is that a WIMP bath that can condensate furnishes the condition to have torsional contributions with a constant of about $10^{8}$ in Planck units as the source of relevant effects at galactic scales, and it does so in such a way that the tangential velocity turns out to be a constant with the measured value for galactic rotation curves: the fact there is no way in which such a tangential velocity might have been any different from constant makes this model more economic than the usual one where the additional hypothesis of DM density distribution with $\frac{1}{r^{2}}$ behaviour had to be necessarily postulated, and the present model is also more predictive because its results are unavoidable while in the standard model the $\frac{1}{r^{2}}$ behaviour is postulated with no reason other than eventually yielding the results that we already knew we should later obtain; both the present and standard model cannot say anything about the actual value of the constant velocity. The difference of the present and standard approaches is that if DM were not have been observed yet then the standard approach would have never been able to predict it while the present approach might have predicted it anyway; the fact that this approach does not predict the actual parameters of the problem should not be surprising since something must be set empirically in any approach to a description of the galactic rotation curves. Another observational difference between the present and standard model is that here the mass of the WIMP is some $34$ orders of magnitude smaller, which is a dramatic discrepancy, but it may not necessarily be possible to test it in accelerators as high particle-production does not necessarily imply large cross-section\footnote{This situation is similar to the one recently met for the Higgs field: the Higgs field has a mass of about $125 \mathrm{GeV}$ while the Top quark has a mass of about $175 \mathrm{GeV}$ which would make us think that discovering the Higgs should have been easier than discovering the Top; it is the fact that the Higgs has more complex decays than the Top what makes the Higgs energetically favoured in its production but overall disfavoured in its detection than the Top itself.} and thus cosmological indirect measures to discriminate these two mass values must be devised. A more serious experimental issue is that the torsional constant with a value of about $10^{8}$ in Planck units has no impact on particle physics and in cosmology we have that the present treatment of Dark Matter constitutes its sole application, while on the other hand it would be desirable to have an alternative physical situation in which to see torsional effects for an independent evaluation of the torsional constant.

On the other hand, Dark Matter is not only supported by galactic rotation curves, but also in terms of other galactic dynamics such as the Bullet Cluster, or more intriguing situations such as Tidal Galaxies, and even circumstances such as galactic formation, CMB and BAO, lying at the interface between cosmology and particle physics: further applications of the hypothesis presented here must also address all of these issues as well; clearly our primary interest here was a first application, so it is beyond the aim of this paper to address them at this stage, but in general what is expected to happen is that every time in which conditions arise for which the WIMP particles can condensate under gravitational pull then the torsionally-induced non-linear dynamics must be manifest. To be more specific is not possible because non-linear behaviours are too sensitive to the environmental conditions of each single specific application.

To conclude, we would like to use the present study as a specific example of a more general problem: as already discussed, in the standard approach torsion is neglected and then additional assumptions have to be postulated; in general, this signifies that we are ready to refuse something we have been given as gift to buy extra assumptions in order to have the means to achieve what we would have accomplished had not we thrown anything away, and this could render lame a model that otherwise might have worked well instead.

We believe this study has shown that too.


\begin{thebibliography}{50}
\bibitem{Clowe:2006eq}
D.~Clowe, M.~Bradac, A.~H.~Gonzalez, M.~Markevitch,\\ 
S.~W.~Randall, C.~Jones, D.~Zaritsky, \textit{Astrophys. J.} \textbf{648}, L109 (2006).
\bibitem{a-g/1}
D.~V.~Ahluwalia, D.~Grumiller,
JCAP \textbf{0507}, 012 (2005).
\bibitem{a-g/2}
D.~V.~Ahluwalia, D.~Grumiller,
\textit{Phys. Rev. D} \textbf{72}, 067701 (2005).
\bibitem{a-l-s/1}
D.~V.~Ahluwalia, C.~Y.~Lee, D.~Schritt,
\textit{Phys. Lett. B} \textbf{687}, 248 (2010).
\bibitem{a-l-s/2}
D.~V.~Ahluwalia, C.~Y.~Lee, D.~Schritt,
\textit{Phys. Rev. D} \textbf{83}, 065017 (2011).
\bibitem{dr-hs}
R.~da Rocha, J.~M.~Hoff da Silva,
\textit{J. Math. Phys.} \textbf{48}, 123517 (2007).
\bibitem{fabbri/1}
L.~Fabbri,
\textit{Mod. Phys. Lett. A} \textbf{25}, 151 (2010).
\bibitem{fabbri/2}
L.~Fabbri,
\textit{Mod. Phys. Lett. A} \textbf{25}, 2483 (2010).
\bibitem{fabbri/3}
L.~Fabbri,
\textit{Gen. Rel. Grav.} \textbf{43}, 1607 (2011).
\bibitem{fabbri}
L.~Fabbri,
\textit{Phys. Lett. B} \textbf{704}, 255 (2011).
\bibitem{h-h-k-n}
F.~W.~Hehl, P.~Von Der Heyde, G.~D.~Kerlick, J.~M.~Nester,\\
\textit{Rev. Mod. Phys.} \textbf{48}, 393 (1976).
\bibitem{f/1a}
L.~Fabbri,
in \textit{Annales de la Fondation de Broglie:\\ Special 
Issue on Torsion} (Ed. Dvoeglazov, Fondation de Broglie, 2007).
\bibitem{f/1b}
L.~Fabbri,
in \textit{Contemporary Fundamental Physics:\\ Einstein 
and Hilbert} (Ed. Dvoeglazov, Nova Science, 2011).
\bibitem{f/2a}
L.~Fabbri,
\textit{Annales Fond. Broglie} \textbf{33}, 365 (2008).
\bibitem{f/2b}
L.~Fabbri,
\textit{Int. J. Theor. Phys.} \textbf{51}, 954 (2012).
\bibitem{f/3}
L.~Fabbri,
\textit{Mod. Phys. Lett. A} \textbf{27}, 1250028 (2012).
\bibitem{Baekler:2011jt}
P.~Baekler, F.~W.~Hehl,
\textit{Class. Quant. Grav.} \textbf{28}, 215017 (2011).
\bibitem{Fabbri:2011kq} 
L.~Fabbri,
arXiv:1108.3046 [gr-qc].
\bibitem{Fabbri:2012ag} 
L.~Fabbri,
arXiv:1210.1146 [gr-qc].
\bibitem{Boehmer:2007um} 
C.~G.~Boehmer, T.~Harko,
\textit{JCAP} \textbf{0706}, 025 (2007).
\bibitem{Silverman:2002qx} 
M.~P.~Silverman, R.~L.~Mallett,
\textit{Gen. Rel. Grav.} \textbf{34}, 633 (2002).
\bibitem{Tilquin:2011bu} 
A.~Tilquin, T.~Schucker,
\textit{Gen. Rel. Grav.} \textbf{43}, 2965 (2011).
\end{thebibliography}
\end{document}